\documentclass[prb,aps,amsmath,preprintnumbers,twocolumn,amssymb]{revtex4}
\usepackage{graphicx}
\usepackage{dcolumn}
\usepackage{bm}
\usepackage[dvips]{color}

\begin{document}

\preprint{to appear in The Journal of Chemical Physics}

\title[Nanoscopic spontaneous motion of liquid trains]
{Nanoscopic spontaneous motion of liquid trains: 
non-equilibrium molecular dynamics simulation}

\author{Amir Houshang Bahrami}
\author{Mir Abbas Jalali}
\email{mjalali@sharif.edu}
\affiliation{Computational Mechanics Laboratory, 
             Department of Mechanical Engineering \\
             Sharif University of Technology, Azadi Avenue, Tehran, Iran}

%%%%%%%%%%%%%%%%%%%%%%%%%%%%%%%%%%%%%%%%%%%%%%%%%%%%%%%%%%%%%%%%%%%%%
%% The document title should be given as usual
%% A short title can be given as a *suggestion* for running headers.
%%%%%%%%%%%%%%%%%%%%%%%%%%%%%%%%%%%%%%%%%%%%%%%%%%%%%%%%%%%%%%%%%%%%%

\date{\today}
% It is always \today, today, but any date may be explicitly specified

\begin{abstract}
Macroscale experiments show that a train of two immiscible 
liquid drops, a bislug, can spontaneously move in a capillary tube because 
of surface tension asymmetries. We use molecular dynamics simulation of 
Lennard-Jones fluids to demonstrate this phenomenon for NVT ensembles in 
sub-micron tubes. We deliberately tune the strength of intermolecular forces 
and control the velocity of bislug in different wetting and viscosity conditions. 
We compute the velocity profile of particles across the tube, and explain 
the origin of deviations from the classical parabolae. We show that the 
self-generated molecular flow resembles the Poiseuille law when the ratio 
of the tube radius to its length is less than a critical value.  
\end{abstract}

%\pacs{47.61.-k, 47.55.-t, 47.15.-x, 47.60.-i, 47.11.Mn}
% PACS, the Physics and Astronomy Classification Scheme.

\maketitle

\section{Introduction}

Dynamics of liquid flow in nanoscopic systems has received great attention 
in recent years. Design and manufacturing of nanofluidic devices, fluid 
transport phenomenon \cite{WHI07,TCC02}, the recent development of 
``lab-on-a-chip'' technology \cite{C06,MEvdB05} and fluid imbibition 
in the nanopores of biomembranes, are amongst notable applications. 
Our knowledge, however, has been mostly confined to the dynamics of 
capillary flows at macroscales \cite{Wash21,Yang04} where the relationships 
between the velocity and contact angle with the surface tension, 
dynamic viscosity and wetting characteristics are known \cite{Gen85,Chan05}.

Generating and controlling the driving force of liquids in extremely 
thin tubes is a major technological challenge, which suggests the 
application of self-propelling systems. The spontaneous motion of 
liquid drops was initially reported by Marangoni \cite{Mar71} and it was 
shown to occur in variably wet and temperature controlled tubes 
\cite{Wei97,MH00}, chemically reacting fluids \cite{deG98,BBM94,DdSO95} 
and in surface-tension-driven motions \cite{Bic00}. 
In all these cases the driving force comes from physical or environmental 
asymmetries, which may cause a net motion. In the present study, we 
are interested in the surface-tension-driven motion of droplets because 
of its importance in microfluidic automation \cite{Chan05,Squ05}. 
In particular, we intend to scale down (by several orders of magnitude)
a self-propelling liquid system, called a bislug, which was introduced 
by Bico \& Qu\'er\'e \cite{Bic00,Bic02}. Although the motion of a bislug 
is fairly explained by the dynamics of Newtonian fluids in macroscales, 
there is neither theoretical nor experimental evidence for the 
applicability/validity of the same physical rules for molecular 
ensembles at sub-micron scales. 

\begin{figure}
\centerline{(a)\hbox{\includegraphics[width=0.4\textwidth]
             {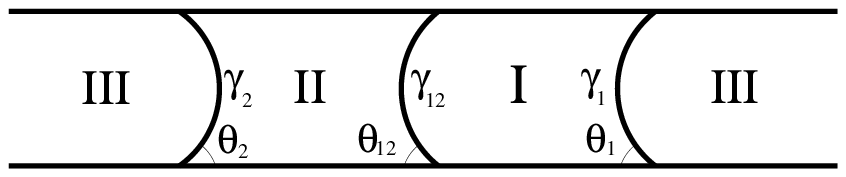}} }
\centerline{(b)\hbox{\includegraphics[width=0.4\textwidth]
             {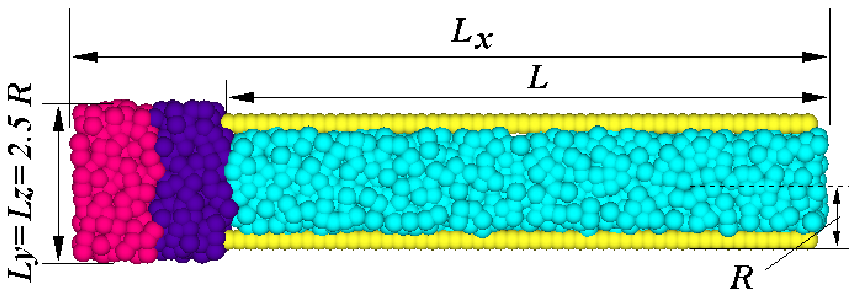}} }
\caption{(a) Configuration of a liquid bislug in a nanotube. 
(b) Initial state of our LJ fluids. Fluid III initially fills 
the tube and liquids I and II occupy the reservoir.}
\label{fig:fig1}
\end{figure}

A liquid bislug is composed of two liquid droplets, liquid I and liquid II,
juxtaposed inside a capillary tube and bounded by a third fluid (fluid III)
on both sides (Fig. \ref{fig:fig1}{\em a}). The system therefore involves 
three interfaces each called a meniscus. The difference between adhesive 
interactions of two neighboring fluids directly determines the meniscus 
wetting power, and so its contact angle. In the partial wetting state,
the contact angle is finite, and it reduces to zero in the limit of 
complete wetting condition. Moreover, the cohesive interactions inside 
each of two adjacent fluids as well as adhesive interactions between 
them, determine the meniscus surface tension. The total driving force 
exerted on a bislug is 
\begin{equation}
F \propto \gamma_d=\gamma_1 \cos\theta_1 + 
\gamma_{12} \cos\theta_{12} - \gamma_2 \cos\theta_2,
\label{eq:resultant-driving-force}
\end{equation}
where $\gamma_d$ is the effective surface tension, and the parameters 
($\gamma_1,\gamma_{12},\gamma_2$) and ($\theta_1,\theta_{12},\theta_2$) 
are, respectively, the surface tensions and contact angles 
(Fig. \ref{fig:fig1}{\em a}). 

In macroscale experiments, when all menisci partially wet the tube wall, 
the contact angles are changed so that the forces at three menisci compensate 
each other and keep the liquid bislug in static equilibrium \cite{Yang04,Bic02}. 
As soon as at least one of the menisci completely wets the tube, its contact 
angle drops to zero and the existing force balance breaks down. If liquid II 
is more wetting than liquid I (see Fig. \ref{fig:fig1}{\em a}), it will be deposited 
on the tube wall while liquid I moves with a constant velocity. In this article, 
we carry out a molecular dynamics simulation of a nanoscopic system of immiscible 
liquids, and reveal its self-propelling dynamics for a wide range of material and 
interfacial properties.

\section{Model Description}
\label{sec:modeling}

We use the setup of Fig. \ref{fig:fig1}{\em b} 
to incrementally turn on the force contribution of each meniscus. 
Using fluid III, we simulate the presence of air or any other fluid and 
equalize its pressure at both sides of the bislug using periodic boundary 
conditions. Fluid III also suppresses possible diffusion effects. We consider 
a cylindrical tube of the length $L$ and radius $R$ composed of fixed solid 
particles that do not interact with each other. The wall particles are 
distributed on a cylindrical arrangement with a constant distance of 
$\sigma$, which is our length scale. This is how we build a rigid impenetrable 
wall. The tube is attached to a $(L_x-L)\times L_y\times L_z(L_y=L_z=2.5R)$ 
boxy reservoir on the left side. The height (and also width) of the reservoir 
is larger than the tube diameter, and the joint between the tube and the reservoir 
is rigid. We apply periodic boundary conditions at the lateral walls of the 
reservoir (perpendicular to the view shown in Fig. \ref{fig:fig1}{\em b}) 
as well as in the longitudinal direction: all liquid particles may freely 
enter the reservoir if they reach to the rightmost boundary of the tube. 

There are four different types of particles corresponding to 
liquid I, liquid II, fluid III, and the solid wall, which are numbered 
in our subsequent formulation as 1, 2, 3, and 4, respectively. Fluid 
particles are located initially at a cubic lattice with the lattice 
constant of $\sigma$, and all particles interact through the Lennard-Jones 
(LJ) potential 
\begin{equation}
U_{ij}(r) = 4\epsilon_{ij} \left [\left ( \frac{\sigma_{ij}}{r} \right )^{12}-
\beta_{ij} \left ( \frac{\sigma_{ij} }{r} \right )^{6} \right ].
\end{equation}
Here $\epsilon_{ij}$ and $\sigma_{ij}$ define the strength and effective range 
of $U_{ij}$ between the particles of types $i$ and $j$. Theoretically, the radius 
of influence of the LJ potential extends to infinity and all binary interactions 
will involve non-zero finite forces. Nevertheless, the computational cost will be 
drastically decreased if one cuts off the attractive LJ force for radii above some 
$r_c$. For all interactions, we choose $r_c=2.5\sigma$, which has already been used 
in the literature \cite{Mar02,DMB07,DMB08} and it guarantees a drop of the attractive 
force below $2\%$ of its maximum value. The choice of $r_c=2.5\sigma$ has also resulted
in a reasonable match between molecular dynamics simulations and experimental 
data \cite{Rao76}. We control the mixing of different fluids by setting \cite{Her05} 
\begin{equation}
\beta_{ij}=
\left \{ 
  \begin{array}{ccc}
    1   & {\rm if} & i=j, \\ 
    0.1 & {\rm if} & i\not = j, \\ 
  \end{array}
\right.  
\end{equation}
for $i,j=1,2,3$. 

The surface tension at each interface depends on the internal 
interactions of each fluid as well as interactions between two 
fluids, controlled by $\epsilon_{ij}$ ($i,j=1,2,3$). The wetting power 
of each liquid depends on adhesive interactions between the liquid 
and tube particles, which is solely controlled by $\epsilon_{i4}$ 
in the LJ potential. The difference between the values of 
$\epsilon_{i4}/\epsilon_{ii}$ for adjacent fluids, determines the 
contact angle of their interface (meniscus) such that the concave 
part faces towards the fluid with lower $\epsilon_{i4}/\epsilon_{ii}$. 
The resultant driving force $F$ is controlled by adjusting the values 
of $\epsilon_{ij}$ and so the contact angles \cite{DMB07,DMB08}. 
Moreover, the viscosity of the $i$th fluid, $\eta_i$, is determined 
by $\epsilon_{ii}$.

We implement non-equilibrium molecular dynamics simulation of an NVT 
ensemble \cite{Allen87} using the velocity Verlet algorithm with the 
integration time step $\delta t = 0.005 t_0$ where 
\begin{equation}
t_0 =\sqrt {\frac{m\sigma^2}{48 k_{\rm B} T} } = \frac{1}{\sqrt{38.4} }.
\end{equation} 
The chosen time step corresponds to $10^{-14}$ seconds for a wide range 
of fluids, and its efficiency has already been confirmed in the study of 
liquid argon \cite{Fre02,Allen87}. We use the dimensionless values of $\sigma = 1$, 
the particle mass $m=1$, and set $k_{\rm B}T = 0.8$. To preserve the ensemble 
temperature at a constant value, a Nos\'e-Hoover thermostat \cite{hoover85} is 
applied at each step. We scale all lengths and velocities by $\sigma$ and 
$\sigma/t_0$, respectively.

\section{Simulation Results}

We do our simulations when: (A) all fluids have the same viscosity 
$\eta$ and surface tension $\gamma$ (B) viscosities of fluids differ 
from each other. We then generate the normalized displacement $x/L_x$ 
of the front meniscus (FM) versus the normalized time $t/t_0$ for 
different values of $\alpha=R/L$. The position of the front meniscus
$x$ is measured from the leftmost boundary of the reservoir, and it
is defined as the location where the density of liquid I reaches to 
$90\%$ of its maximum value. Although the experiments of type A are 
unrealistic, they help us to independently investigate the influence 
of contact angles on the bislug dynamics. We study three possible 
circumstances: (i) only the rear meniscus (between liquid II and 
fluid III) completely wets the tube (ii) the rear and middle menisci 
completely wet the tube (iii) all three menisci completely wet the tube. 
We use two tubes of radii $R=10.34$ and $5.17$.

\subsection{Equal Viscosities}

We set $\epsilon_{11}$=$\epsilon_{22}$=$\epsilon_{33}$=1 to create equal viscosities,
and use $\epsilon_{34}=0.5$ and $\epsilon_{12}$=$\epsilon_{23}$=$\epsilon_{13}$=1
to get the same boundary interactions between different fluids. The internal 
interactions of each fluid is governed by $\beta_{ii}$=1 ($i$=1,2,3) and we set 
$\beta_{12}$=$\beta_{23}$=$\beta_{13}$=0.1 \cite{Her05} to prevent liquids 
from mixing. We start our computations with a case that all menisci partially 
wet the tube. The contact angles in a partial wetting condition lie in the range 
$0<\theta_1,\theta_{12},\theta_2<90^{\circ}$ should one choose the values of 
$0.5\leq\epsilon_{14},\epsilon_{24}\leq1.1$. In such a circumstance, the capillary 
force pulls both liquids into the tube subsequently, while fluid III leaves the 
tube and enters the reservoir. 

\begin{figure}
%\centerline{(a) \hspace{3in} (b)}
%\vspace{0.2in}
\centerline{\hbox{\includegraphics[width=0.42\textwidth]
             {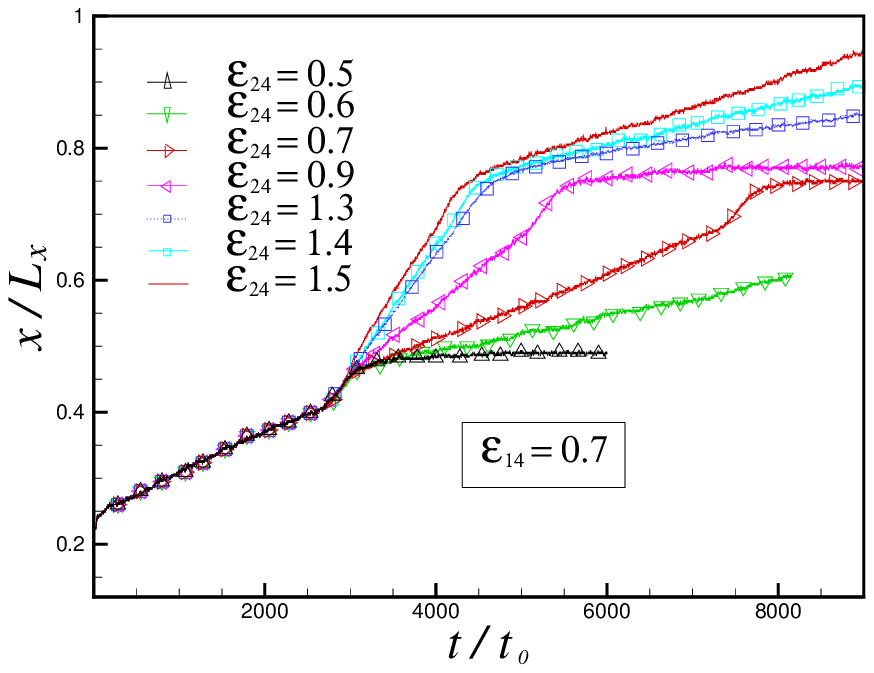}}            
           }
\caption{The normalized displacement curve of the front meniscus 
($x/L_x$) versus time in partial and complete wetting conditions 
with $\alpha=0.107(R=10.34,L=96.14)$ and $\eta_{1}=\eta_{2}=\eta_{3}$.}
\label{fig:fig2}
\end{figure}

As each meniscus enters the tube, the driving force is altered and 
the slope of the displacement curve of the FM varies accordingly. 
One can therefore identify three phases on displacement curves 
(Fig. \ref{fig:fig2}). The first phase begins when liquid I 
enters the tube and it ends when liquid II flows into the tube. 
The driving force of the first phase is proportional to 
$\gamma_d=\gamma \cos\theta_1$. The second phase starts by the 
inflow of liquid II and continues until the last molecules of 
liquid II leave the reservoir. This stage involves two active 
menisci implying $\gamma_d=\gamma(\cos\theta_1+\cos\theta_{12})$. 
In the third (and last) phase, the whole bislug is inside the tube 
and the driving force is computed from Eq. (\ref{eq:resultant-driving-force}). 

Fig. \ref{fig:fig2} displays the displacement history of the FM 
for several values of $0.5 \leq \epsilon_{24}\leq 0.9$ and for 
$\epsilon_{14}=0.7$. By keeping $\epsilon_{14}$ constant, we 
have fixed the driving force of the FM. For the three cases 
$\epsilon_{24}$=$0.5$, $0.6$, and $0.7$ that satisfy 
$\epsilon_{24} \le \epsilon_{14}=0.7$, the concave side of the 
middle meniscus (between liquids I and II) faces liquid II 
($\theta_{12}\geq90^{\circ}$) and it resists the driving force 
of FM. Consequently, in the second phase the bislug moves slower 
than the first one. For $\epsilon_{24}=0.5$ we have $\epsilon_{24}$=
$\epsilon_{34}$, whose natural implication is $\epsilon_{14}-\epsilon_{24}$=
$\epsilon_{14}-\epsilon_{34}$, which in turn leads to the condition 
$\theta_{12}+\theta_{1}=180^{\circ}$. This says that the surface 
tension of the front meniscus cancels that of the middle one, and 
the bislug must stop as soon as liquid I leaves the reservoir 
(zero velocity in the second part of displacement curve in 
Fig. \ref{fig:fig2}{\em a}). By increasing $\epsilon_{24}$, the velocity 
of FM increases as long as liquid II has not completed its entrance 
into the tube. However, after the complete drainage of liquid II 
from the reservoir, the displacement curve of the FM becomes flat 
indicating a static equilibrium in partial wetting conditions: 
$\gamma(\cos\theta_1 + \cos\theta_{12} - \cos\theta_2)=0$.

\begin{figure}
%\centerline{(a) \hspace{3in} (b)}
%\vspace{0.2in}
\centerline{\hbox{\includegraphics[width=0.42\textwidth]
             {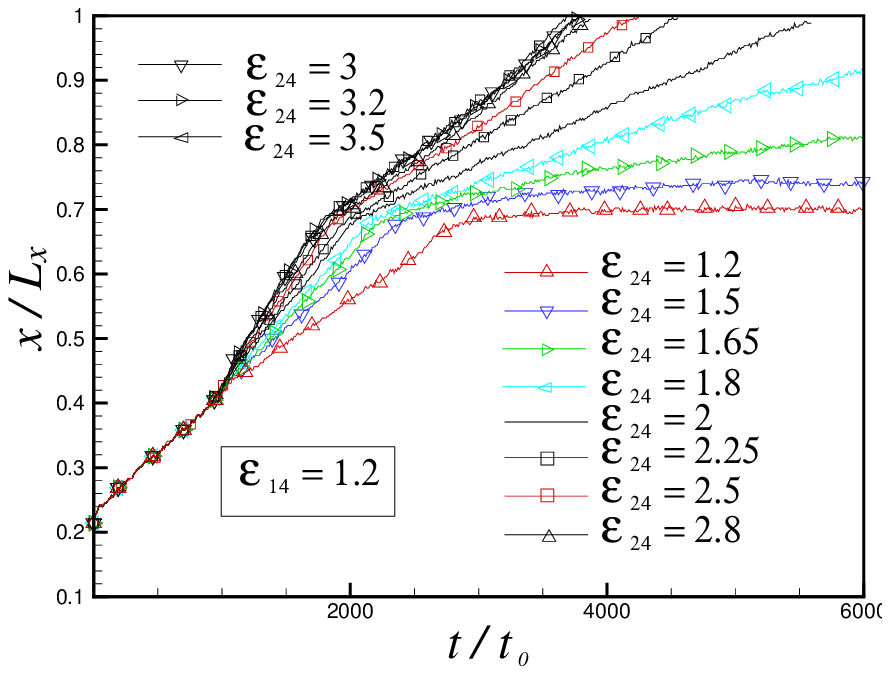}}            
           }
\caption{The normalized displacement curve of the front meniscus 
($x/L_x$) versus time in partial and complete wetting conditions 
with $\alpha=0.107$ ($R$=10.34, $L$=96.14) and $\eta_{1}=\eta_{2}=\eta_{3}$.}
\label{fig:fig3}
\end{figure}

\begin{figure}
\centerline{\hbox{\includegraphics[width=0.4\textwidth]
             {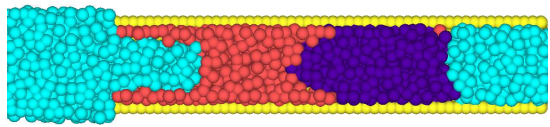}} }
\caption{\label{fig:fig4} A moving bislug corresponding to 
the case of $\epsilon_{14}$=$0.7$ and $\epsilon_{24}$=1.5 
in Fig. \ref{fig:fig2}.}
\end{figure}

For $\epsilon_{24} \ge 1.2$ and with the same $\epsilon_{14}=0.7$, 
the rear meniscus completely wets the tube, and results in 
$\theta_2=0$ and $F\propto\gamma_d=\gamma(\cos\theta_1 + \cos\theta_{12} - 1)$. 
This is the first complete wetting situation, which holds over the 
range $1.2 \le \epsilon_{24} < 1.5$. Since $\cos\theta_1+\cos\theta_{12} >1$, 
the bislug undergoes a sustained motion after entering the tube until 
liquid II is totally consumed, or the FM reaches the end of the tube
(Fig. \ref{fig:fig2}). By further increasing $\epsilon_{24}$, we obtain 
a critical value of $\epsilon_{24}\approx 1.5$ beyond which the middle 
meniscus completely wets the tube. This is the second complete wetting 
state ($\theta_{2}=\theta_{12}=0$) and the force components corresponding 
to $\gamma\cos\theta_2$ and $\gamma\cos\theta_{12}$ cancel each other 
identically. The bislug is thus driven by the FM: 
$F\propto\gamma_d=\gamma\cos\theta_1$. Fig. \ref{fig:fig4} shows a snapshot 
of the moving bislug in its third phase. It is evident that liquid II 
leaves a thin stationary layer on the tube wall. 

To obtain the state of complete wetting condition for all three menisci, 
we fix $\epsilon_{14}=1.2$ and carry out new simulations (see Fig. \ref{fig:fig3}). 
For the very special case of $\epsilon_{14}=\epsilon_{24}=1.2$ that implies 
$\theta_{1}=\theta_{2}=0$ and $\theta_{12}=90^{\circ}$, and according to 
our assumption $\gamma$=$\gamma_1$=$\gamma_{12}$=$\gamma_2$, liquids I 
and II will behave as a single droplet, which is pulled into the tube by 
the maximum force $F_{\rm I}\propto \gamma$ exerted by the FM, but it 
stops once the rear meniscus enters the tube and contributes 
$F_{\rm II}\propto -\gamma$ to the total driving force. We then increase 
$\epsilon_{24}$ so that the force components $\gamma \cos\theta_1$ and 
$\gamma \cos\theta_2$ cancel each other and the driving force comes from 
the middle meniscus whose concave side faces liquid I for 
$\epsilon_{24}>\epsilon_{14}$. The middle meniscus, however, will be in 
partial wetting as long as $\cos \theta_{12}<1$. Therefore, the velocity 
of the FM (and so the bislug) is boosted by increasing $\epsilon_{24}$ 
because $\cos \theta_{12}\propto \epsilon_{24}-\epsilon_{14}$. 
The four last graphs of Fig. \ref{fig:fig3} show that the velocity of 
the FM is saturated for $\epsilon_{24} > 2.8$, which corresponds to the 
complete wetting condition of all menisci. 

The experiments with $(\epsilon_{14},\epsilon_{24})=0.7$ in Fig. \ref{fig:fig2} 
and $(\epsilon_{14},\epsilon_{24})=1.2$ in Fig. \ref{fig:fig3} show that the 
bislug velocity in the second phase is slightly less than the first phase while 
the equality of $\epsilon_{14}$ and $\epsilon_{24}$ should have resulted in  
$\theta_{12}=90^{\circ}$ and identical velocities during the first and second phases
of motion. Our explanation of this phenomenon is as follows: while the middle meniscus
is driven, it is subject to back-and-forth oscillations that alternately change 
the contact angle and keep its average magnitude larger than the expected value
(this is because the particles near the contact point of a meniscus usually lag 
the field particles, which feel less frictional force). For the special case of 
$\epsilon_{14}=\epsilon_{24}$, the average value of $\theta_{12}$ marginally 
exceeds $90^{\circ}$ and causes a tiny resistive force. Therefore, the velocity 
of the bislug is slightly reduced during the second phase. So far we have 
considered different cases of bislug motion when three fluids have the same 
viscosity $\eta$. Below we vary the viscosities of fluids and investigate realistic 
cases.

\subsection{Different Viscosities}

We use Kirkwood-Buff's expression   
\begin{equation}
\gamma=\int dx \left \{ p_{xx}(x)-\frac 12 \left [
p_{yy}(x)+p_{zz}(x) \right ] \right \},
\end{equation}
and compute the surface tension $\gamma$ from the elements of pressure 
tensor using the equilibrium molecular dynamics simulation of two adjacent,
thermostated liquids in a cube \cite{Kirk49}. In similar conditions, we apply 
the Green-Kubo integral to obtain the viscosity of our LJ fluids \cite{Allen87,RW82}.
The element $p_{xx}(x)$ of the pressure tensor is normal to the interface. 
For the system of fluids with equal viscosities and surface tensions, we find 
$\gamma=2.6\pm 0.15$ and $\eta=3.25 \pm 0.25$. It is remarked that the application 
of other methods such as shear flow, gives compatible results \cite{BLHL05}.

\begin{figure}
\centerline{\hbox{\includegraphics[width=0.42\textwidth]
             {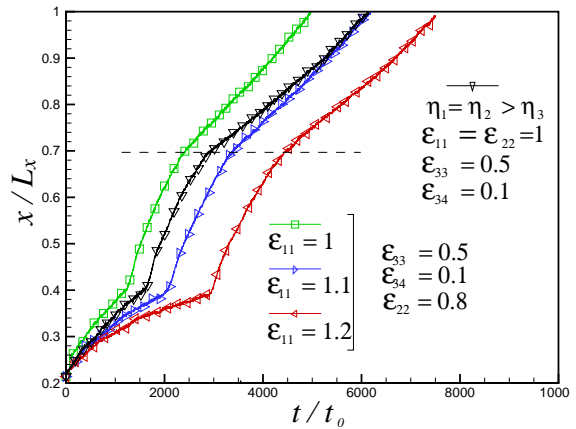}}  }
\caption{The normalized displacement curve of the front 
meniscus ($x/L_x$) versus time for $\alpha=0.107$ ($R$=10.34, $L$=96.14), 
$(\epsilon_{14},\epsilon_{24})=(0.7,2.3)$ and different viscosities 
$\eta_{1}>\eta_{2}>\eta_{3}$. Here, the FM is in partial wetting 
condition and the other two menisci completely wet the tube wall. 
We demonstrate the special case of $\eta_{1}$=$\eta_{2}>\eta_{3}$ 
to show the effect of $\eta_1$ and $\eta_2$ on the bislug velocity 
independently. Horizontal dashed line approximately indicates the 
beginning of the third phase.}
\label{fig:fig5}
\end{figure}

We now vary the viscosities of fluids around $\eta \approx 3.25$. Increasing 
the value of $\epsilon_{ii}$ enables us to increase the viscosity of the 
$i$th fluid, $\eta_i$. By taking $\epsilon_{34}=0.1$ and 
$\epsilon_{ii}$=$(0.5,0.8,1,1.1,1.2)$, we find
\begin{equation}
\eta_{i}=(1.6\!\pm \!0.1,2.6\!\pm \!0.2,
3.25\!\pm \!0.25,3.6\!\pm \!0.32,4.\!\pm \!0.33),
\end{equation}
and apply our NVT code to compute $x(t)/L_x$ when
$\eta_{1}>3.25>\eta_{2}>\eta_{3}=1.6$. Some of our results 
are displayed in Fig. \ref{fig:fig5}. As one could anticipate, 
the first two phases of displacement curves are no longer linear 
versus time, and the profile of the first phase resembles a 
parabolic pattern $x/L_x\propto\sqrt{t/t_0}$, which is the well-known
Washburn equation \cite{DMB07}. It is evident that even for the case 
of different viscosities, the third phase of each displacement curve 
has still a constant velocity in accordance with macroscale experiments 
\cite{Bic00}. 

By comparing the results of Figures \ref{fig:fig2} and \ref{fig:fig5} 
we conclude that reducing $\eta_3$ (with respect to $\eta_1$ 
and $\eta_2$) boosts the bislug velocity considerably. 
It is remarked that we needed to decrease $\epsilon_{34}$ while
we were doing so for $\epsilon_{33}$ just to preserve the wetting 
condition of the FM. In Fig. \ref{fig:fig5}, comparing the graphs of 
$(\epsilon_{11},\epsilon_{22})=(1,0.8)$ and
$(\epsilon_{11},\epsilon_{22})=(1,1)$ suggests that decreasing   
$\eta_2$ (while $\eta_1$ is kept constant) induces an even faster 
motion because the deposition of liquid II on the tube wall 
speeds up. The graphs corresponding to $\epsilon_{22}=0.8$ 
show that the bislug velocity is slowed down by increasing $\epsilon_{11}$, 
which means higher values of $\eta_1$ and therefore a larger
viscous force. This effect is clearly evident from the longer first 
phase (similar to $\sqrt{t/t_0}$) of the bislug displacement curve.

\section{Physics of the Flow}

\begin{figure}
\centerline{\hbox{\includegraphics[width=0.4\textwidth]
             {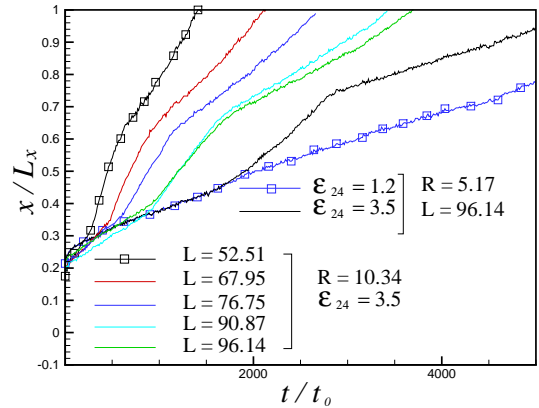}} }
\caption{The normalized displacement curve of the front meniscus ($x/L_x$) 
versus $t/t_0$ for $0.107\le \alpha \le 0.197$ with $R=10.34$, and for 
$\alpha=0.053$ with $R=5.17$. All menisci are in complete wetting condition, 
and we have set $\epsilon_{14}=1.2$ and $\eta_{1}=\eta_{2}>\eta_{3}$.}
\label{fig:fig6}
\end{figure}

In all cases, the velocity of the FM, $V(t)=dx/dt$, is constant over 
the last phase of displacement curves, and it is proportional to 
$\gamma_d$. The motion with a constant $V$ is due to the balance 
between the driving force $F=2\pi R\gamma_d$ and the viscous friction 
force $F_{\rm drag}$ from the tube wall. i.e., $F-F_{\rm drag}=m dV/dt=0$ 
where $m$ is the total mass inside the tube. When all fluids have the 
same viscosity $\eta$ and the flow is Poiseuille, one obtains 
\begin{equation}
V\equiv V_p=\frac{\gamma_d\alpha}{4\eta},~~
F_{\rm drag}=8\pi \eta L V_p.
\label{eq:two}
\end{equation}
In this equation, $\eta$  and $L$ must be replaced, respectively,
with $\eta_1=\eta_2$ and the bislug length when fluid III has 
a negligible contribution to the dynamics. Given $\alpha=R/L$, 
the magnitude of $V_p$ can therefore be calculated. To see in which 
conditions the bislug velocity attains the value of $V_p$, we carry out
some new calculations with the two radii $R=5.17$ and $10.34$, and by 
taking several tube lengths of $52.51\le L \le 96.14$, which yield 
$0.107 \le \alpha \le 0.197$. We use fluids of the same viscosity, 
set $\epsilon_{14}=1.2$, and let all menisci completely wet the tube wall. 
Some of our results have been illustrated in Fig. \ref{fig:fig6}. 
According to our simulations, the quantity $V/L_x$ tends to $V_p/L_x$ by 
decreasing $\alpha$ and the relative error $|V-V_p|/V_p$ reduces from $33\%$ 
(for $\alpha=0.197$) to $3\%$ (for $\alpha=0.107$). In the complete wetting 
condition of all menisci and for $\epsilon_{24}=3.5$, we find the critical 
value $\alpha_{\rm cr}\approx 0.1$ below which the velocity of the FM becomes 
identical to $V_p$. A couple of our experiments correspond to $R=5.17$ and 
$L=96.14$ that give $\alpha=0.053$. With such a small ratio and for
$\epsilon_{14}=\epsilon_{24}=1.2$, liquids I and II remain separated but they 
behave as a single drop. For the stronger wall interactions of liquid II with 
$\epsilon_{24}=3.5$, the relative velocity error further decreases and we 
obtain $V\approx V_p$. This is evident from Fig. \ref{fig:fig6} as the first 
and third phases of motion have the same slope. 

We are using a Nos\'e-Hoover thermostat, which is mainly applied 
to equilibrium phenomena. The self-propelling motion of liquid trains, however,
is a non-equilibrium case and a thermostat may destroy the Galilean invariance of 
dynamical equations. Through computing the Washburn prefactor, we show that such 
an effect is quite small in most of our simulations that correspond to narrow 
tubes of $\alpha \le \alpha_{cr}$. When fluids I and III are immiscible and have 
the same viscosity, the time dependence of the displacement function $x/L_x$ is 
linear during the entrance of liquid I into the tube, and the Washburn equation 
$V\sim t^{-1/2}$ is transformed to the ordinary law of capillary dynamics: 
$V=\gamma \alpha \cos \theta/(4\eta)$. This becomes $V=0.2\alpha$ for 
($\gamma,\eta$)=($2.6,3.25$) in our setup and in the complete wetting condition 
of the FM ($\theta=\theta_{1}=0$). For $\epsilon_{14}=1.2$ and $\alpha=0.107$, the 
first phase of the displacement curves in Fig. \ref{fig:fig3} corresponds to the 
complete wetting of the FM and the slope of this phase gives $V \approx 0.0225$. 
On the other hand, the capillary equation yields $V=0.2\times 0.107=0.0214$. 
For the case with $(R,\alpha)=(5.17,0.053)$ and $\epsilon_{14}=\epsilon_{24}=1.2$ 
in Fig. \ref{fig:fig6}, the capillary equation leads to $V=0.2\times 0.053=0.0106$ 
while the slope of the first phase of the displacement curve is $V\approx 0.0108$. 
The relative errors between theoretical predictions and our simulations are therefore 
$5\%$ and $1.8\%$ for $\alpha=0.107$ and $\alpha=0.053$, respectively. This result 
shows that the effect of the Nos\'e-Hoover thermostat on the Galilean invariance 
of governing equations is insignificant, specially when $\alpha<\alpha_{cr}$.

\begin{figure}
\centerline{\hbox{\includegraphics[width=0.4\textwidth]
             {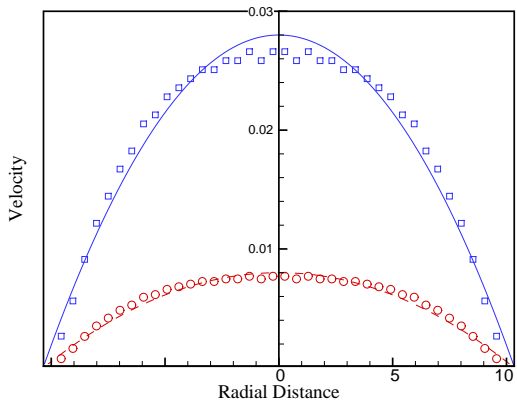}} }
\caption{The profile of $\bar v(r)$ versus the radial distance $r$ from the tube 
centerline. Circles correspond to the case of equal viscosities with 
$(\epsilon_{14},\epsilon_{24})=(0.7,1.5)$ in Fig. \ref{fig:fig2}.
Squares show the profile of the case with $(\epsilon_{11},\epsilon_{22})=(1,0.8)$ 
in Fig. \ref{fig:fig5} (for different viscosities). Solid and dashed lines are 
the best-fitted parabolic curves.}
\label{fig:fig7}
\end{figure}

Let $v(\zeta,r,t)$ denote the longitudinal velocity of molecules inside the 
tube where $r$ and $\zeta$ are the distances of molecules from the tube 
centerline and the reservoir outlet, respectively, and $t$ is the time. 
Small circles and squares in Fig. \ref{fig:fig7} display the simulated 
profile of 
\begin{equation}
\bar v(r)= \lim_{t \rightarrow \infty} \frac 1t \int_{0}^{t}
dt  \frac {1}{L} \int_{L_x-L}^{L_x} v(\zeta,r,t) d\zeta,
\label{eq:averaged-v}
\end{equation}
together with the best-fitted parabolic curves to the data. These data have 
been obtained for the case with $(\epsilon_{14},\epsilon_{24})$=$(0.7,1.5)$ 
in Fig. \ref{fig:fig2}, and $(\epsilon_{11},\epsilon_{22})$=$(1,0.8)$ in 
Fig. \ref{fig:fig5}. The time average in equation (\ref{eq:averaged-v}) 
is computed only over the third phase of bislug motion, and averaging 
along the tube length helps us to reduce the noise of dynamic capillary waves 
and work with a better statistical data. It is seen that for $\alpha=0.107$, 
the flow inside the tube is almost Newtonian and laminar. There is a better 
match between the simulated data and the best-fitted parabolic curve near 
the wall because the boundary layer (due to shearing) initially develops 
near the wall and then penetrates into inner regions. In other words, 
molecules near the tube centerline are subject to radial motions more than 
those living near the wall. Our numerical experiments show that the 
functional form of $\bar v(r)$ does not depend on the contact angle 
$\theta_{12}$ but on $\alpha$. By increasing $\alpha$, the gradient 
$\frac{d}{dr}\bar v(r)$ increases near the wall, enhances $F_{\rm drag}$ 
and we get $V<V_p$. The existence of $\alpha_{\rm cr}$ has also been 
confirmed by the macroscale experiments of Bico \& Qu\'er\'e \cite{Bic02}. 
They report $\alpha_{\rm cr} \approx 0.07$ and ascribe the drop of 
velocity for $\alpha > \alpha_{\rm cr}$ to the viscous dissipation at 
the menisci. Below, we argue that the physical mechanism of the process 
is different in molecular ensembles.

The information of any developing phenomenon, like meniscus oscillation 
and boundary layer formation, is transmitted by individual molecules that 
interact by their neighbors and after several ``kicks" lose their initial 
memory. The oscillations of driving menisci remain unseen for downstream 
particles when the tube length is increased: for a long bislug, the number 
of longitudinal interactions between molecules is quite large and the coming 
information from oscillating menisci is gradually dissolved. What actually 
matters in the radial direction is the development of the boundary layer 
whose propagation into central regions is again through interactions between 
molecules. When the tube radius is small, the number of radially disturbing 
kicks that molecules experience is small and the boundary layer can fully 
extend to the tube centerline. In tubes of larger radii, however, molecules 
undergo too many radial interactions (kicks) and lose the ``shearing information" 
that they had to carry to molecules near the tube centerline. As a result, 
the boundary layer is not fully developed and some amount of the kinetic 
energy is carried by radial motions. An evidence for this process is the 
flattening of the profile of $\bar v(r)$ in central regions (see Fig. \ref{fig:fig7}). 
To have a Poiseuille flow in sub-micron scales, we thus need both a small 
radius (measured in units of $\sigma$) and a large tube length. These would 
mean the existence of some $\alpha_{\rm cr}$, which should correlate with 
the relaxation time of the bislug.

\section{Conclusions}

The bislug motion is a consequence of the transfer of free energy,
from initially separated liquids, to the longitudinal component of 
the kinetic energy of molecules. The system chooses 
a definite direction of motion because the different wetting 
capabilities of liquids I and II breaks the symmetry along the 
tube axis. Independent of the viscosities of fluids, we observed a 
constant bislug velocity when both liquids I and II leave the reservoir. 
The Washburn law governs the history of $x(t/t_0)/L_x$ in the 
entrance phase of liquid I to the tube \cite{DMB07}. The bislug 
will not move (the first four cases in Fig. \ref{fig:fig2}) when 
the adhesion with the wall is not enough for making a permanent 
and stable binding between all molecules of a component (either 
of liquids I and II) and the tube wall.  

We showed that the spontaneous motion of immiscible liquids 
inside the tubes of sub-micron radii is an efficient process in certain 
wetting conditions, and it leads to significant mass transport should one 
use appropriate liquids. An external pressure gradient has played a decisive 
role in previous experiments \cite{M05TM09} and simulations \cite{TMc09} of 
water flow in carbon nanotubes. Our fundamental achievement was to capture 
the flow characteristics of a nanoscopic self-propelling system in the 
absence of an external pressure field. The size of the bislug, which 
indeed determines the relaxation time, and its relation with the form 
of the velocity profile across a small tube, helped us to suggest a new 
physical mechanism for explaining the deviations from the Poiseuille law. 
Our results have potential applications in mass transport in fluidic 
devices and also in the separation of immiscible liquids in sub-micron 
scales.  

\section{Acknowledgments}
We thank the referees for their constructive comments, which helped us 
to improve the presentation of our simulations. This work was partially 
supported by the research vice-presidency at Sharif University of Technology.

%\bibliography{ahb}% Produces the bibliography via BibTeX.

\end{document}